\documentstyle[twocolumn,aps]{revtex}
\begin{document}
\twocolumn[\hsize\textwidth\columnwidth\hsize\csname
@twocolumnfalse\endcsname
\title
{\begin{flushright}
{\normalsize CERN-TH/2000-248}\\
\vspace*{0.4 cm}
\end{flushright}
On the two kinds of vector particles}
\author
{M. V. Chizhov}
\address
{Theory Division, CERN, CH-1211 Geneva 23, Switzerland\\
and\\
Centre for Space Research and Technologies,
Faculty of Physics,\\ University of Sofia, 1164 Sofia, Bulgaria}
\maketitle

\vspace{.5cm}
\hfuzz=25pt
\begin{abstract}
All known elementary vector particles, 
the photon, $Z$, $W$ and the gluons,
are described by the gauge theory. They belong to the {\it real}
representation $(1/2,1/2)$ of the Lorentz group. On the other hand
{\it inequivalent} representations $(1,0)$ and $(0,1)$ also correspond
to particles with spin 1. It is natural to suppose that, along with
the known vector particles, the new particles can exist.
Evidence for the existence of these particles in nature is the presence
of the axial-vector meson resonances with quantum numbers $1^{+-}$. Other
indications for their existence are discussed. The
signatures of their contributions into different physical processes are
presented.
\end{abstract}
\vskip2pc]

\section{Introduction}
The description of the elementary particles relies on concepts of 
symmetry~\cite{nov}.
The rotation invariance group gives a particle classification 
with respect to the spin. In the quantum theory
the spin may have half-integer or integer value. The lowest
representation of the $O(3)$ group,
which can be used as a building block for the construction of a
higher spin representation, is a two-component spinor $\psi_\alpha$
($\alpha=1,2$). It describes the particles with spin 1/2. 
In non-relativistic quantum mechanics there exists only one
possibility to construct a vector: {\bf 1/2~+~1/2~=~1}. 

In the relativistic theory the symmetry group is
the Lorentz group $O(3,1)$,
which is isomorphic to the direct product of the two spatial rotation
groups $O(3)\times O(3)$. Therefore, two inequivalent representations
$(1/2,0)$ and $(0,1/2)$ exist for the spin 1/2. 
They correspond to particles with different chiralities
and are represented by the Weyl spinor 
$\psi_\alpha$ and
its conjugate $\psi^*_\alpha\equiv \psi_{\dot\alpha}$,
which are related by the $P$ transformation.
There exist two possibilities to construct the representation
of the spin 1.

The vector representation $(1/2,1/2)$ is the well studied one and 
corresponds to the gauge particles. It is chirally neutral,
because it arises from the product between the left $(1/2,0)$
and the right $(0,1/2)$ fundamental spinors. 
This property reflects the simple fact
that all gauge interactions preserve chiralities of incoming and
outgoing particles. This representation
is transformed as a mixed (dotted and 
undotted) spinor $\phi_{\alpha\dot\beta}$, which is equivalent
to the Lorentz vector
\begin{equation}
\label{vector} 
V_m=(\sigma_m)^{\dot\alpha\beta}\phi_{\beta\dot\alpha},
\end{equation}
where $\sigma_i$ $(i=1,2,3)$ are the Pauli matrices and
$\sigma_0$ is the unit matrix.

Let us stress that {\it inequivalent chiral} representations $(1,0)$ and
$(0,1)$, which also correspond to particles with spin 1, 
can be constructed if one uses only the product either
of the left $(1/2,0)$ or of  
the right $(0,1/2)$ fundamental spinors. They are transformed
independently by the proper Lorentz group as rank-2 spinors,
symmetric in the spinor indices: $\phi_{\alpha\beta}$ and
$\phi_{\dot\alpha\dot\beta}$, respectively. To pass to more convenient
Lorentz indices, the decomposition of the product of the Pauli matrices
into symmetric and antisymmetric parts can be used:
\begin{mathletters}
\begin{eqnarray}
&&(C\hat\sigma_m\sigma_n)^{\alpha\beta}=
g_{mn}C^{\alpha\beta}-
\frac{i}{2}~\epsilon_{mnab}(C\hat\sigma^a\sigma^b)^{\alpha\beta},
\\
&&(\sigma_m\hat\sigma_n C)^{\dot\alpha\dot\beta}=
g_{mn}C^{\dot\alpha\dot\beta}+
\frac{i}{2}~\epsilon_{mnab}(\sigma^a\hat\sigma^b C)^{\dot\alpha\dot\beta},
\end{eqnarray}
where
$(\hat\sigma_m)_{\alpha\dot\beta}=(C^{-1}\sigma_m^T C)_{\alpha\dot\beta}$,
$C_{\alpha\beta}\equiv\epsilon_{\alpha\beta}$ is the charge-conjugation
matrix, and $\epsilon^{mnab}$ is the completely antisymmetric tensor,
with $\epsilon^{0123}=+1$. Therefore, one can introduce the 
antisymmetric anti-self-dual tensor 
$T^-_{mn}=T_{mn}-\tilde T_{mn}$:
\end{mathletters} 
\begin{mathletters}
\label{tensor}
\begin{equation}
T^-_{mn}=\frac{i}{2}~\epsilon_{mnab}(C\hat\sigma^a\sigma^b)^{\alpha\beta}
\phi_{\beta\alpha},
\end{equation}
and the antisymmetric self-dual tensor 
$T^+_{mn}=(T^-_{mn})^*=T_{mn}+\tilde T_{mn}$:
\begin{equation}
T^+_{mn}=
-\frac{i}{2}~\epsilon_{mnab}(\sigma^a\hat\sigma^b C)^{\dot\alpha\dot\beta}
\phi_{\dot\beta\dot\alpha},
\end{equation}   
where $\tilde T_{mn}=(i/2)\epsilon_{mnab}T^{ab}$ is the tensor that is
dual to the real antisymmetric tensor $T_{mn}$.
Hence, the real tensor $T_{mn}$ corresponds both to the left and to
the right vector particles or, in other words,
to the vector and axial-vector ones. 
\end{mathletters}

As far as the vector potential $V_m$ (\ref{vector}) describes the real
vector particles, the antisymmetric tensors $T^{\pm}_{mn}$ (\ref{tensor})
can correspond to yet unknown vector particles. 
It is clear that the transformation law
of the vector particles uniquely defines their 
trilinear renormalizable interaction 
with the fermions without derivatives.
For example, the usual gauge interaction term
$\overline{\Psi}\gamma^m\Psi~V_m$
arises from the Lorentz-invariant form
$\psi^\alpha\phi_{\alpha\dot\beta}\psi^{\dot\beta}$, where
\begin{eqnarray}
\Psi=\left(
\begin{array}{c}
\psi_\alpha\\
iC^{\dot\alpha\dot\beta}\psi_{\dot\beta}
\end{array} \right)=\left(
\begin{array}{c}
\Psi_L\\
\Psi_R
\end{array} \right),~~
\gamma^m=\left(
\begin{array}{cc}
0 & \hat\sigma^m\\
\sigma^m & 0
\end{array} \right),
\nonumber
\end{eqnarray}
are the Dirac bispinor and the Dirac matrices in the helicity
representation. On the other hand, the trilinear interaction of the new
vector particles with the fermions
$\psi^\alpha\phi_{\alpha\beta}\psi^\beta+
\psi^{\dot\alpha}\phi_{\dot\alpha\dot\beta}\psi^{\dot\beta}$ corresponds
to the Yukawa term\footnote{For the gauge antisymmetric tensor field
$B_{mn}$ the gauge-invariant form of the interaction with derivative
$\epsilon^{abcm}\partial_{[a}B_{bc]}\overline{\Psi}\gamma_m\Psi$
is used, which is non-renormalizable.
Here we will consider the antisymmetric tensor matter field $T_{mn}$
with a Yukawa interaction.}
$\overline{\Psi}\sigma^{mn}\Psi~T_{mn}$, where 
$\sigma^{mn}=i~\left[\gamma^m,\gamma^n\right]/2$.
The key feature of the
interactions mediated by the new vector particles is chirality flip of
incoming and outgoing particles.
The gauge vector particles and the new ones have different
interactions and, consequently, different
quantum numbers.  Therefore, they correspond to different kinds of
vector particles. A common
opinion exists in the literature, that massive vector particles can be
equivalently described by the vector potential $V_m$ or by the
antisymmetric tensor field $T_{mn}$~\cite{cyr}. This is true for free
particles as far as both representations correspond to particles with spin 
1. However, there is no equivalence when an interaction is included.
Only interactions, in particular with fermions, can distinguish 
between these two kinds of vector particles. 
To reveal this let us consider the hadron vector resonances.

\section{Hadron vector resonances}
The bound states of a quark and an antiquark are characterized by
the quantum numbers $J^{PC}$, where $J$ is the total angular momentum,
$P$ is the parity and $C$ is the charge conjugation. There exist three
types of different quantum numbers for the known vector mesons~\cite{PDG}. 
They are $1^{--}$, $1^{++}$ and $1^{+-}$. For instance, the first 
quantum numbers are assigned to the $\rho$, $\omega$ and $\phi$ 
vector mesons. The second quantum numbers are assigned to
the $a_1$ and $f_1$ axial-vector mesons; however, note that
the third quantum numbers are assigned again to the
axial-vector mesons $b_1$ and $h_1$. The key point is the
difference between the last two assignments for the axial-vector mesons.

Let us consider the extended
Nambu--Jona-Lasinio (ENJL)~\cite{NJL} models.
In such models the Lagrangian contains only the quark fields, while
the spontaneous symmetry breaking and the hadron states
are generated dynamically by the model itself. The mesonic states
arise as excitations of quark--antiquark pairs and that defines their
interactions with the quarks. There are the vector 
$\overline{\Psi}\gamma^m\Psi$ and the axial-vector
$\overline{\Psi}\gamma^m\gamma^5\Psi$ bilinear forms
of the quark spinor fields
with quantum numbers $1^{--}$ and $1^{++}$,
which correspond to the vector and axial-vector mesons, 
respectively. They have gauge-like interactions with the quarks
and can be described by the gauge vector $V_m$ and 
axial-vector $A_m$ fields.
Up to now all local ENJL models include only vector mesons with
quantum numbers $1^{--}$ and $1^{++}$ and do not describe
mesons with quantum numbers $1^{+-}$. 
One way of incorporating these mesons into
the ENJL model has been described in~\cite{TNJL}.
These mesonic states correspond to the vector particles, which are
described by the antisymmetric tensor field $T_{mn}$, rather than by the
gauge fields. The bilinear form $\overline{\Psi}\sigma^{mn}\Psi$
is used to describe the quantum numbers and the interactions of these
mesons. 
The existence of the axial-vector mesons with the quantum numbers $1^{+-}$
points out that the new kind of vector particles should exist
in nature. 

What concerns the hadron physics the introduction of the
antisymmetric tensor field in the ENJL model can
give new understanding of the vector mesonic resonances.
The six components of the antisymmetric tensor
correspond to the axial-vector 
$B_m=\partial_n(\overline{\Psi}\sigma^{mn}\gamma^5\Psi)$
with quantum numbers $1^{+-}$ and also to
the vector $R_m=\partial_n(\overline{\Psi}\sigma^{mn}\Psi)$
with quantum numbers $1^{--}$.
Each of these vectors $B_m$ and $R_m$
has three independent components
due to the antisymmetric property of $\sigma^{mn}$.
Therefore, besides the axial-vector mesons $B_m$ 
with quantum numbers $1^{+-}$, there are additional vector mesons $R_m$
with quantum numbers $1^{--}$, like those of the
gauge mesons $V_m$, but having different coupling to the quarks.
As far as there exist two different vector particles with the same
quantum numbers $1^{--}$, their mass eigenstates can be a linear 
combination of them\footnote{It means, in particular, that the $\rho$
meson can have both gauge and anomalous tensor couplings with the quarks,
while the axial-vector mesons with quantum numbers $1^{++}$
have only gauge interactions and 
the axial-vector mesons with quantum numbers $1^{+-}$
have only tensor interactions.}.
For example, for the isospin~1 vector mesons
they are $\rho(770)$ and $\rho'$, where the latter is either
the $\rho(1450)$ or $\rho(1700)$ state. For the axial-vector mesons
it is possible to make a unique identification of the low-lying mesonic
states with the quantum numbers $1^{++}$ and $1^{+-}$ as
$a_1(1260)$ and $b_1(1235)$, respectively.
Applying the ENJL approach to all these (axial)-vector mesons,
a remarkable relation among their masses is derived~\cite{TNJL}:
\begin{equation}
m^2_{\rho}+m^2_{\rho'}=m^2_{a_1}+m^2_{b_1}.
\label{mass}
\end{equation}
The extraction of the $a_1$ and $\rho'$ masses from the experiments is
a model-dependent procedure. The masses have a broad range of values.
Therefore, to fix one of these masses one can use as approximation
the Weinberg sum relation~\cite{W}
\begin{equation}
m_{a_1}\simeq\sqrt{2}~m_\rho.
\end{equation}
This fixes the $a_1$ mass value $m_{a_1}\simeq 1088$ MeV,
and favours the choice of $\rho(1450)$ resonance
as the second mass eigenstate with the quantum numbers $1^{--}$,
because eq.~(\ref{mass}) gives $m_{\rho'}\simeq 1450$ MeV.
Certainly, the new point of view on the hadron (axial)-vector
resonances enables us not only to explain their mass spectrum,
but also introduces a new decay dynamics connecting with their
tensor couplings to the quarks.

\section{Electroweak physics}
Since two types of vector particles are used 
for the hadron resonance description, one can extrapolate this 
feature to the electroweak physics. Let us introduce a new type
of intermediate vector bosons into the Standard Model (SM).
As far as the tensor interaction does not conserve chirality, 
these bosons should couple to 
the left doublets and to the right singlets of the fermion
fields. In order to have $SU(2)\times U(1)$ invariance, the new
vector bosons must be doublets. Therefore, they do not mix with
the gauge bosons before the symmetry breaking.
It means that in the high energy experiments they do not interfere
with the gauge bosons and contribute to processes
as their squared amplitude.

Nevertheless, due to the particular chiral properties of the new
particles, the angular distribution of the fermion annihilation
cross-section for the tensor coupling differs from the one of
the vector coupling.
For the vector case at high energies the well-known
expression holds\footnote{For simplicity
the parity violation and the interference between
the photon and $Z$ in the SM can be neglected.}
\begin{mathletters}
\begin{equation}
\label{V}
{{\rm d}\sigma\over{\rm d}\cos\theta}\sim 
\left(1+\cos^2\theta\right).
\end{equation}
While the anomalous tensor interaction
of the currents $\partial_n(\overline{\Psi}\sigma^{mn}\Psi)$
leads to a different angular dependence
\begin{equation}
\label{T}
{{\rm d}\sigma\over{\rm d}\cos\theta}\sim\cos^2\theta.
\end{equation}
\end{mathletters}
The differential cross-section (\ref{T})
also differs from the case
of a scalar coupling, which has uniform distribution.

This means that if the new kind of vector particles exists,
at high energies they should give
an essential contribution to the
differential cross-section only near to the beam
direction. It is not easy to detect definitely
such an excess experimentally, but 
it will be a clear signature of the new kind of
interactions. Some indications of such excesses exist
in recent experimental data with high pseudorapidity $\eta \ge 2$:
in the production of the $b$-quarks at the Tevatron~\cite{Tevatron}
and the $\tau\bar\tau$ pairs production at LEP~2~\cite{LEP}.

The LEP experiments at CERN are very suitable for a search of
such deviations from the SM. Generally, the lepton physics is free from
uncertainties connected with parton distributions, which are 
typical for the hadron colliders. As far as the new interactions
do not conserve chirality, their effects increase with the
mass of the particles. 
Therefore, the analysis of heavy particle decays, such a $\tau$ leptons,
gives a unique possibility for searching these effects.
It was noted in \cite{muon}
that the new interactions lead to another parametrization of pure
leptonic decays, different from the conventional Michel
parametrization. The experimental analysis of the $\tau$ decays
and constraints on the new 
tensor coupling are presented in \cite{DELPHI}.

The low energy physics is able
either to constrain or to indicate the presence
of the new kind intermediate particles. 
This can be realized mainly in precision
experiments of particle decays. However, at present they 
do not obtain the deserved attention.
It is noteworthy to point out here that,
for example, the principal $\rho$ parameter in
the muon decay has not been measured since 1969. Needless to say
that a knowledge of the precision spectrum shape
is important for the determination of the Fermi coupling constant $G_F$.

If some new interactions give a contribution into the muon decay, this
will effectively lead to a lower value of $G_F$ in comparison with
the experimentally extracted one, which assumes only SM interactions.
It may help to solve the long-standing problem of 
a violation of unitarity in the first row of 
the Cabibbo--Kobayashi--Maskawa (CKM) matrix~\cite{Jaus}:
\begin{equation}
\left| V_{ud}\right|^2+\left| V_{us}\right|^2+\left| V_{ub}\right|^2=
0.9968\pm0.0014.
\label{CKM}
\end{equation} 
The main contribution into the unitarity sum (\ref{CKM})
comes from the $V_{ud}$ element of the CKM matrix. This element 
is extracted  with high precision
from the nuclear superallowed 
beta decays $0^+ \to 0^+$, comparing the strength of this vector   
transition to $G_F$:
\begin{equation}
\left| V_{ud}\right|^2 = \mbox{const}/G_F^2.
\end{equation}
A two per mille lowered  value of $G_F$ will restore the 
unitarity condition.

In principle, the new tensor interactions can be responsible for such a
scenario~\cite{muon}. In this case the strength of these interactions must
be comparable with the electroweak one and that can lead to additional
observable contributions in other experiments. 
Let us point out some indication for a possible admixture
of such interactions in 
radiative pion decay~\cite{Bolotov} and semileptonic 
three-body kaon decay~\cite{Akimenko}.

First of all it is important to note that the tensor interactions
do not contribute directly to the pion decay $\pi \to e\nu$
and avoid the strong constraints on the pseudoscalar
interactions. Although the tensor interactions, like the scalar ones,
do not conserve chirality and should give a large contribution into
chirality-suppressed pion decay, this is impossible because
a tensor form factor for the pion matrix element 
$\langle0|\bar{q}\sigma^{mn} q|\pi(p)\rangle$ cannot be constructed. 
A different situation arises
when the radiative pion decay $\pi \to e\nu\gamma$ is considered.
In this case the tensor interactions may give a contribution themselves
and, moreover, may interfere with 
a QED inner bremsstrahlung contribution.
Exactly such a type of destructive interference was observed in the
radiative pion decays in flight~\cite{Bolotov}.

The other experiment, where the tensor form factor can be introduced
on the same footing as the vector form factor of the SM, is the
semileptonic three-body
kaon decay $K \to \pi e \nu$. As far as most of the experiments
on kaon decays give information only about the vector
form factors, the experimental data are poor. However, the last 
high-statistics experiment~\cite{Akimenko} 
indicates the presence of non-zero
tensor form factor. 

In ref.~\cite{pik} it was shown that both the
anomalous destructive interference in the radiative pion decay and
the presence of a non-zero tensor form factor 
in the semileptonic kaon decay,
may be explained by an admixture of tensor interactions. This is also
compatible with the contribution of the tensor interactions into the
pure leptonic decays at the per mille level~\cite{muon}.
It is clear that new experiments are necessary to confirm or reject
the presence of this new type of vector particles in nature.\\

\begin{center}
{\bf ACKNOWLEDGEMENT}
\end{center}

I would like to thank Igor Boiko for the fruitful discussions.
I am also grateful to A. Dolgov and D. Kirilova
for the careful reading of this manuscript and the valuable comments.
I appreciate the warm hospitality of the Theory Division at CERN,
where this work has been done.

\pagebreak[2]

\end{document}